# A Commodity SBC-Edge Cluster for Smart Cities

Basit Qureshi*, Kamal Kawlaq*, Anis Koubaa†*, Basel Sultan*, Mohammad Younis‡
*Department of Computer Sc., Prince Sultan University, Saudi Arabia
†CISTER/INESC TEC and ISEP-IPP, Portugal
‡Department of Computer Sc. & Elec. Engg., University of Maryland, Baltimore, MD 21250, USA
Email: *{qureshi,214110623,akoubaa,215110432}@psu.edu.sa; ‡younis@cs.umbc.edu

*Abstract:* The commodity Single Board Computers (SBCs) are increasingly becoming powerful and can execute standard operating systems and mainstream workloads. In the context of cloud-based smart city applications, SBCs can be utilized as Edge computing devices reducing the network communication. In this paper, we investigate the design and implementation of a SBC based edge cluster (SBC-EC) framework for a smart parking application. Since SBCs are resource constrained devices, we devise a container-based framework for a lighter foot-print. Kubernetes was used as an orchestration tool to orchestrate various containers in the framework. To validate our approach, we implemented a proof-of-concept of the SBC based Edge cluster for a smart parking application, as a possible IoT use-case. Our implementation shows that, the use of SBC devices at the edge of a cloud based smart parking application is a cost effective and low energy, green computing solution. The proposed framework can be extended to similar cloud-based applications in the context of a smart city.

## 1. Introduction

In recent times, Smart City initiatives adopted by governments has led corporations, institutions and municipal bodies to adopt smart technologies in order to enhance citizens living standards. Cities are already using smart traffic lights, smart transportation, smart buildings, smart waste collection, smart infrastructure management etc. It is estimated that by 2022, there will be more than 19 billion Internet of Things (IoT) devices connected to the Internet [1].

Typically, cloud computing technology with Internet connectivity enables smart cities, where sensors / smart things collect information and relay it to distantly located data centers for compute and storage [2]. Cloud based smart city systems rely on centralized architecture where data sources such as sensors transmit information to a centralized data center for storage and processing [3]. An advantage of this setup is that all data is stored and processed in a single location making data management and post processing simpler. A disadvantage is that large amounts of data needs to be transmitted from data sources to the data center increasing the network traffic therefore constraining the network bandwidth resulting in latency. A key challenge for this centralized architecture is the seamless network connectivity and reliable cloud service for smart applications to work.

Edge computing [4] [5], [6] [7] as a technology provides a reliable aspect of hosting smart city applications. Instead of hosting compute / storage in a data center (centralized servers), a set of smaller clusters / cloudlets collaborate locally to process and store information closer to the data sources. Consequently, the communications overhead and latency is significantly reduced by adopting a de-centralized architecture within a smart city enabling system. This demonstrates that edge computing satisfies the need to decentralize the storage and compute capabilities of smart city systems.

Single Board Computers (SBC) have now become sufficiently powerful to run standard operating systems including Microsoft Windows 10, Linux distributions such as Ubuntu, Debian etc. Raspberry Pi [8] SBC has sold in much higher volumes earning hefty profits to the manufacturers. This has led towards a dramatic rise of SBC manufacturers developing SBCs in various form factors with increasingly powerful devices at lower costs. Each of these competing devices has been subject to different design criterion with a large variation in functionalities available. The low price offered by SBCs is an attractive opportunity for creating SBC based clusters. C. Baun [9] developed a small Raspberry Pi based cluster with 5 SBCs for educational purposes; they conduct extensive performance analysis of the cluster from various perspectives. Parallel Computing sculpture [10] provides an array of 256 Raspberry Pi computers used for sculpturing at Virginia Tech. Erica the Rhino [11] uses 5 Raspberry Pi computers to allows users to interact with an art project. Qureshi et. al., [12] developed a cluster for image analysis in a robotics environment using SBCs. They deploy Hadoop on this cluster and evaluate performance characteristics of the cluster with various big data workloads. To the best of our knowledge, there is a lack of motivation in utilizing SBC based clusters as edge computing devices certainly due to the availability of limited on-board processing and memory resources.

In this paper, we investigate the use of SBC-edge clusters (SBC-EC) as a disruptive technology to enable edge computing in smart cities systems. SBC-EC clusters provide an opportunity to extend the cloud computing paradigm to the edge where a micro cloud cluster exists between the data-center and the data source at the edge in a smart city application deployment. The data source or IoT devices may consist of sensors and actuators that run on batteries and usually without a strong processor. This leads to the necessity of offloading computation to nearby SBC cluster rather than a distant data center. Generally, the network latency, quality of service in transmission and computation / service time are important to achieve higher dependability. The SBC-EC clusters process intermediate data and later transmit the processed information to the data center for further processing or storage. Consequently, the network traffic is reduced, affecting the traffic congestion, therefore improving the network latency and throughput, which are key characteristics for seamless operability of a smart city application. Fig 1 depicts the concept of using SBC-EC clusters as edge clusters for a smart city management system. In the context of, i) smart street lighting, ii) smart waste management and iii) smart home; sensors relay information to a SBC-EC cluster periodically. The cluster processes relevant information on-site, and later transmits the processed information to the data center; reducing the overall network traffic. The contribution of this paper is in two folds:

- We present a framework utilizing SBC based clusters on the edge in a smart city environment using Kubernetes and Docker container technologies. The light-weight containers are suitable for a SBC based cluster due to their light foot-print.
- We develop and evaluate a working prototype for a smart parking application that periodically reads data from the parking sensors. The data is processed on the SBC cluster and later transmitted to the cloud.

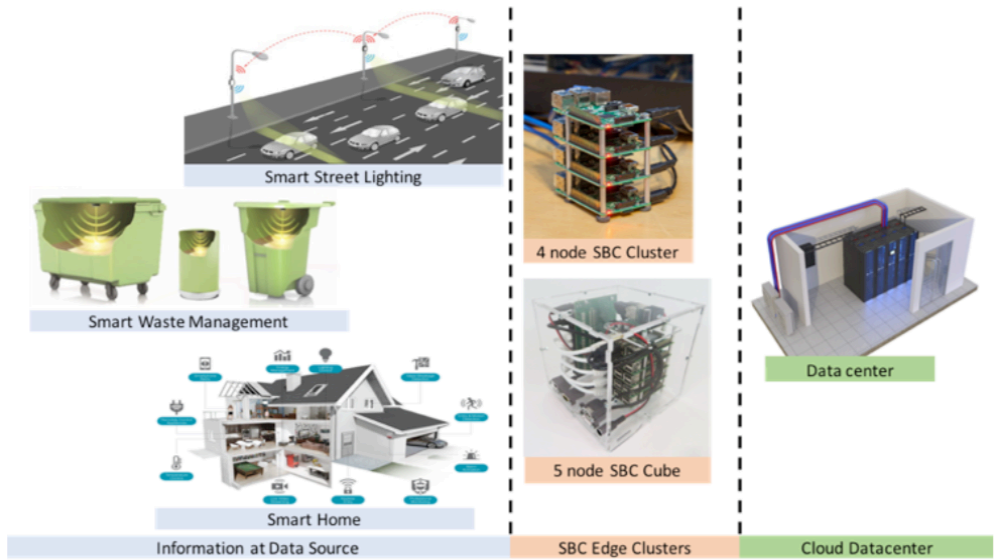

Fig. 1: SBC Edge cluster in a smart city context

TABLE I: Comparison of SBC Platforms

| SBC Platform | Processing Hardware | RAM (GB) | Cost (USD) | Features |
|---|---|---|---|---|
| RASPBERRY PI 3B+ | Broadcom BCM2837B0 | 1 | 35 | HDMI, Composite Video, Gigabit Ethernet, Wi-Fi, Bluetooth, MicroSD, USB 2.0, USB 3.0, I2C, I2S, GPIO. |
| ODROID XU4 | Samsung Exynos5422 | 2 LPDDR3 | 59 | HDMI, USB 2.0, USB 3.0, Gigabit Ethernet, CSI, I2S, I2C, GPIO, eMMC/SDCard up to 64 GB |
| LATTEPANDA 4G | Intel Atom Cherry Z8350 | 4 DD3L | 149 | HDMI, Ethernet, Wi-Fi, Bluetooth, GPIO, microSD up to 64GB. Runs MS-Windows 10. |

The rest of the paper is presented as follows; section 2 presents three popular SBC platforms with detailed comparison of various features of each platform. Section 3 provides details for SBC-EC framework consisting of the SBC-EC cluster design and deployment in the context of a smart parking application. Section 4 presents related works followed by conclusions in section 5.

## 2. Single Board Computer Platforms

Due to the widespread use of smart-phones, manufacturers have entered the SBC market making various chipsets avail- able at low prices. This recent trend has enabled SBC manufacturers to develop various devices with different features and functionalities at lower costs. In this section, we discuss 3 different SBC platforms and provide a comparison based on Processing power, on-board memory, price and various features available on-board.

The Raspberry Pi [8] is predominantly the market leader with a low cost and affordable price at 35 USD (as of July 2018). The device uses a Broadcom BCM2837B0 System- on-chip(SoC) with a quad core processor running at 1.4 GHz. The device allows various modes of connectivity with a Gigabit Ethernet, HDMI, USB, Wi-Fi and Bluetooth. Operating systems such as Ubuntu, Android etc. are supported for the ARM32 based architecture. The Raspberry Pi foundation also provides a light weight Raspbian Operating system.

The Ordoid Xu-4 [13] was developed by HardKernel and is available for 59USD. Offering open source support, the board can execute various flavors of Linux, such as Ubuntu, Ubuntu MATE, Android Oreo. Xu-4 uses Samsung Exynos5 Quad-core ARM Cortex-A15 Quad 2Ghz and Cortex-A7 Quad 1.3GHz CPUs with 2Gbyte LPDDR3 RAM at 933MHz. The Mali-T628 MP6 GPU supports OpenGL 3.0 with 1080p resolution via standard HDMI connector. Two USB 3.0 ports, as well as a USB 2.0 port, allows faster communication with attached devices.

TABLE II: Comparison of SBC Platforms

| System-on-chip (SOC) | Cores/Clock-speed | Graphical Processing Unit(GPU) |
|---|---|---|
| BROADCOM BCM2837B0 | 4 / 1.4 GHz | Broadcom Video Core IV |
| INTEL ATOM CHERRY Z8350 | 4 / 1.8 GHz | Intel HD Graphics 200-500 |
| SAMSUNG EXYNOS5422 | 4 / Cortex-A15 at 2.0GHz and 4 / Cortex-A7 at 1.4GHz | Mali-T628 / MP6 |

The Latte Panda 2G/4G [14] was developed as a SBC device with Microsoft Windows support. This device comes in two different memory capacities with 2GB and 4GB RAM. Both device share the same Intel Atom Cherry Z8350 SoC with a quad core processor at 1.8 GHz supporting Intel HD Graphics 200-500 GPU. This device also supports HDMI, Ethernet, Wi- Fi, Bluetooth connectivity. A micro SD Card with a capacity of up to 64GB can also be installed. The device supports multiple operating systems including Ubuntu, MacOS and OpenSUSE in addition to Microsoft Windows IoT and Windows 10. Table I shows a comparison of the three SBC Platforms. Table II shows a comparison of SoC features on these SBC platforms.

Recent developments in SBCs have enabled the inclusion of more powerful peripherals on the boards. Latte Panda Alpha and Up Squared board are two examples of such SBC, both having Intel processor, large on board RAM and better GPUs. These features currently comes at a higher purchase price (399 USD), but in due time, the cost of these devices might trickle down into lower-cost SBCs. In this work, we build SBC based edge clusters using the Raspberry Pi 3B+, Odroid Xu-4 and LattePanda 4G SBCs. Further details are presented in the next section.

## 3. SBC-EC Framework

In this section we present the details of the SBC-EC framework. In a typical cloud based smart-city application, IoT devices or sensors collect / harvest information and later transmit these to the cloud which is essentially hosted in a data center. The sheer number of sensors transferring data periodically can result in massive network traffic, consequently affecting network latency and creating congestion issues. To address this issue, we propose a framework, that deploys SBC based clusters as edge devices in the cloud. Due to their small form factor and low costs, SBC based edge clusters can be easily and cost effectively deployed on a site. The sensors data can be relayed to the SBC-EC for local processing instead of the distantly located data-center, therefore reducing the need for massive network communication.

The SBC-EC's ecosystem leverages the use of light weight cloud computing technologies suitable for resource constraint SBC devices. The framework utilizes Docker based containers [13], [15] Containers are designed to be lightweight having a small foot print in comparison to the traditional virtual machines (VM). Much of the operating system functions such as file system, memory address space and networking can be virtualized in a container. The small size of containers as opposed to a VM makes it easier to copy and move the container images between devices.

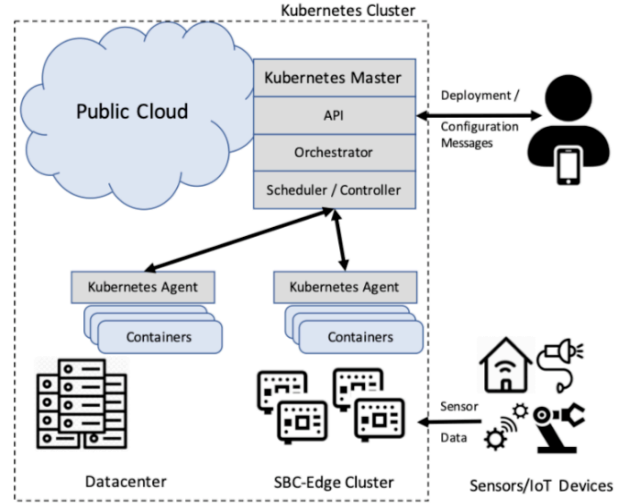

Fig. 2: Kubernetes Eco-System with SBC based Edge-Cluster

Kubernetes [16] is a container orchestration tool which allows different physical machines to host containers. The distributed environment makes it possible to deploy multiple containers executing various instances of an application on multiple physical machines. In the Kubernetes environment, a master node provides a singular point of control providing access control and communication with the slave nodes referred to as Kubernetes agents. A Kubernetes agent is deployed on a physical machine and is responsible for managing the various containers deployed on that physical machine. The Master node is responsible for orchestration which consists of various activities including resource allocation, scheduling, container deployment etc. In the context of a cloud based smart-city application, various containers can be deployed on physical nodes, within the data center and on the edge in the SBC- EC. The Kubernetes orchestrator allows seamless control of all containers regardless of the physical deployment on the physical nodes. Fig 2 presents a high-level architecture of the SBC- EC Framework. We deployed a prototype of this framework at the Prince Sultan University Campus. The framework consists of three main components, i) Smart parking sensors, ii) SBC- EC edge cluster, iii) Smart parking application deployed as a RESTful web-service on containers in the Kubernetes based cluster. Next, we present details for these components.

### A. Smart parking sensors

The physical infrastructure of the framework consists of parking sensors installed in a parking lot. The parking sensors communicate with a local relay that transmits the information from the sensors to the cloud based smart parking applications. In this work, we allocated 22 parking bays at one of the parking lots at the Prince Sultan University campus as can be seen in figure 3. A wireless parking sensor was installed on each parking bay. Each wireless sensor connects to a gateway which is installed within a distance of 30 meters from the parking sensors. The parking sensors have an inbuilt battery and communicate to the gateway using Wi-Fi. The SBC-edge device is connected to the gateway through an API. The smart- parking application executing on the cluster, constantly ping the gateway API periodically with an interval of 60 seconds. The gateway serves as a relay and connects to the SBC-Edge cluster.

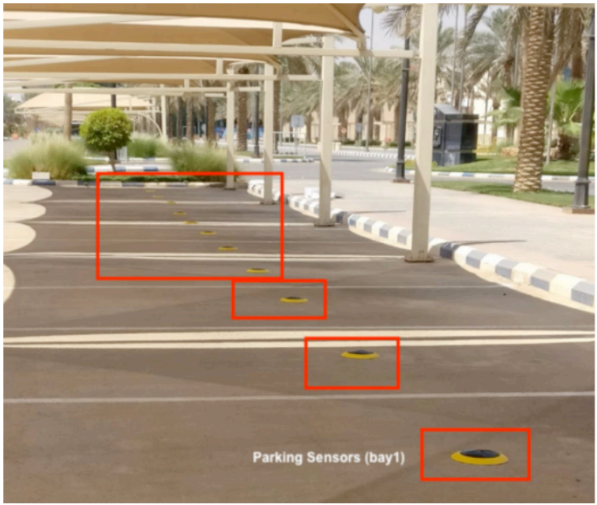

Fig. 3: Parking sensors at the smart-parking-lot at the PSU campus

### B. SBC-Edge Cluster

For testing purposes, we constructed three SBC based edge clusters. Fig 4 shows three SBC-EC clusters. Each SBC- EC cluster is composed of 2-3 SBC devices including a Raspberry Pi, Odroid Xu-4 and Latte Panda. Each of these devices requires a SD-card for storage. Raspbian Operating system was installed on each SBC. Furthermore, we initialized Kubernetes agents/slaves on each device which connects to a Kubernetes master node. The devices in each cluster connect to a local Ethernet switch which is connected to the university network providing Internet access. To realize the cloud based architecture, we installed the Kubernetes master node on a PC in the lab. This PC connects to the SBC-EC clusters over the Internet. The total cost of the SBC-EC clusters was US$ 117, 183 and 304 for Raspberry Pi, Odroid Xu-4 and Latte Panda 4G clusters.

### C. Parking Application

We develop a smart city parking application using the SBC- EC framework. The application deployment on the cluster serves as a prototype implementation of the framework. The main objective of this application is to log data collected from parking sensors, process the data locally in the SBC-EC (total occupation time of parking bays) and send updates to the cloud service.

The application listens to the `gateway` using `SocketIO` for any updates from within a container. After the initial handshake, the application listens to the bays event which sends the current status of the parking bays. All received data is appended with a time stamp, the results are further stored within the container as log files synchronized by the time stamp. Timestamps are used to calculate the occupation time at the end of the specified period. Listing 1 shows the `bays` script used in the application.

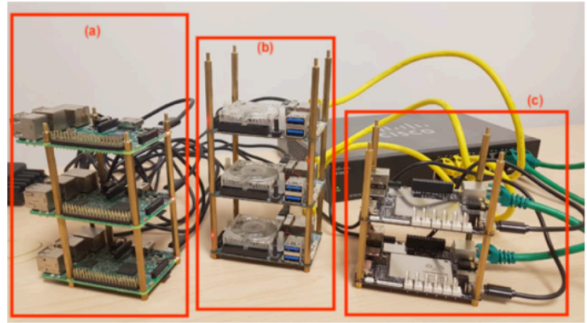

Fig. 4: A Kubernetes agent installed on 3 SBC-EC Clusters. (a) Raspberry Pi (b) Odroid Xu-4 (c) Latte Panda; devices in each cluster connect to a ethernet switch

```
socket.on('bays', (receivedData) => {
  console.log(data);
  receivedData.data.forEach(parkingLot => {
    parkingLot.bays.forEach(bay =>{
      bay['timestamp'] = new Date().getTime();
      bay['occupationTime'] = 0;
      baysList.push(bay);
    });
  });
});
```

Listing 1: `bays` event

Another event is `baysUpdate` which gets triggered at any time the `status` of the parking bay changes. As an update is triggered, the gateway sends relevant information about the selected parking bay. If the status of the parking bay changed from `occupied` to `free`, the difference between the current time-stamp and the stored time stamp is computed, the result is stored in the log file. Similarly, if the status changed from free to occupied, the log file is subsequently updated. Every 24 hours, the application reads the log file to process the data. It determines the occupancy of each parking bay by processing the timestamps of the records annotated as `occupied` within the log file. The results are processed into a Comma Separated Value (CSV) format and can be processed further in the cloud by application managers. Listing 2 shows the `processData` function of the application. The `processData` function is called at regular intervals which can be specified by the user.

### D. Experimental observation

The framework was successfully deployed at the PSU campus. Fig 5 shows the results from our deployment. The SBC-EC cluster was deployed in the parking lot and results recorded for the 3rd week of November 2018. Fig 5 (a) shows the occupancy of all parking bays on a given day of this week. In this instance, the parking occupancy rate averaged 7.5 hours per day. The maximum occupancy rate was observed on Tuesday due to large traffic at campus as a result of the evening exams. Fig 5 (b) shows the average of the minimum and maximum hours occupancy of each parking bay per day, during the week. Two cars were left overnight in parking bays 12 and 21, to determine the accurate functioning of the application. These results demonstrate the accurate functioning of the prototype, validating the proposed framework.

```
function processData() {
  updateOccupationTime();
  const records = [];
  baysList.forEach(bay => {
    records.push({bayID: bay.id, occupationTime:
        .occupationTime, occupationRate: bay.
        occupationRate});
    bay.occupationTime = 0;
  });
  csvWriter.writeRecords(records)
    .then(() => {
      console.log('Done');
    });
}
```

Listing 2: `processData` function

## 4. Related Work

Lim et. al. [17] defined the need for processing large amounts of data in smart cities. They argue that policy makers in the smart cities need to address infrastructure level changes for smart city enabling technologies. The need for cloud and edge based IoT platforms for smart cities to facilitate various services and applications is imminent.

Weisong Shi et. al. [2] presented a vision and discuss the challenges of the Edge Computing paradigm. Based on their study they deduce that increasingly, services are pushed from the cloud to the edge of the network. This is because process- ing data at the edge can ensure shorter response time and better reliability. This paper provides a vision for future research in edge computing. Kochovski in [18] presented the design and architecture of smart IoT construction environments. The proposed three tier architecture provides virtualization and communication mechanism at three levels namely embedded systems, edge gateway and virtual clusters. A video conferencing application executes on the cloud allowing multiple parties at a construction cite to communicate. Sensors information from construction site is shared using the virtualized Kubernetes clusters deployed in the cloud. As compared to our work, this work does not contribute towards SBC based clusters at the edge. Tran et. al. [6] discussed the need for video processing on the edge in the future 5G networks. They present three use-cases that benefit the communication in 5G networks. They conclude that edge computing enables a capillary distribution of cloud computing capabilities to the edge of the network. Researchers in [19] introduced a Lightweight Edge Gateway for the Internet of Things (LEGIoT) architecture. It leverages the container-based virtualization using SBC devices to support various IoT application protocols. They deploy the proposed architecture on a SBC device to demonstrate functioning of an IoT gateway.

Rathore et. al. [20] proposed a combined IoT-based system for smart city development and urban planning using Big Data analytics. A four-tier architecture is proposed that includes IoT sources and data generation; communication; data management and processing using a Hadoop framework, and data analysis layers. They used Apache Spark to process data from various smart cities IoT datasets. The proposed system focuses on centralized processing of information as opposed to exploiting the benefits of computing at the edge. Johnstone et. al [21] presented a review of current SBC technologies. They present a discussion on the various implications of deployment of such devices in Internet of Things and Smart city systems. Morabito [4] provided a performance evaluation study of popular SBC platforms. They compare various SBCs using Docker container virtualization in terms of CPU, Memory, Disk I/O, Network performance criterion. Authors in [22] conducted an extensive performance evaluation on various embedded microprocessors systems including Kinkit Smart and Raspberry Pi. They deploy Docker based containers on the systems and analyze the performance of CPU, Memory and Network communication on the devices. They conclude that the container-based virtualization has a negligible impact in terms of CPU and memory consumption compared to the native (non-virtualized) scenarios.

The above-mentioned works establish the need for a light weight virtualization infrastructure and technologies to support smart services on the edge in a smart city. This work addresses the gap in literature by leveraging the use of low-cost SBC clusters as edge devices to provide cloud-based smart city services using light weight virtualization technologies.

## 5. Conclusions

In this paper we presented the design of a SBC-EC framework for smart parking in the context of a cloud-based smart city application. We deployed Docker containers and Kubernetes agents on the resource constrained SBC clusters. A web-service application was developed to process data from parking sensors which is processed locally in the SBC-EC cluster on the edge. The results are later transmitted to the cloud for further processing and decision making. A prototype of the framework was successfully deployed at Prince Sultan University campus. The SBC-ECs are comparatively inexpensive clusters and can be easily replaced in the event of damage or physical loss. Furthermore, the image-size of the containers executing on the SBC-EC are light-weight, and can be easily copied or moved. The successful deployment of the prototype demonstrates the low cost benefit of deploying such cluster in smart city applications. In future we intend to address the reliability and security issues in the developed prototype. We further intend to study the role of SBC-EC cluster in reducing the power consumption in the data centers.

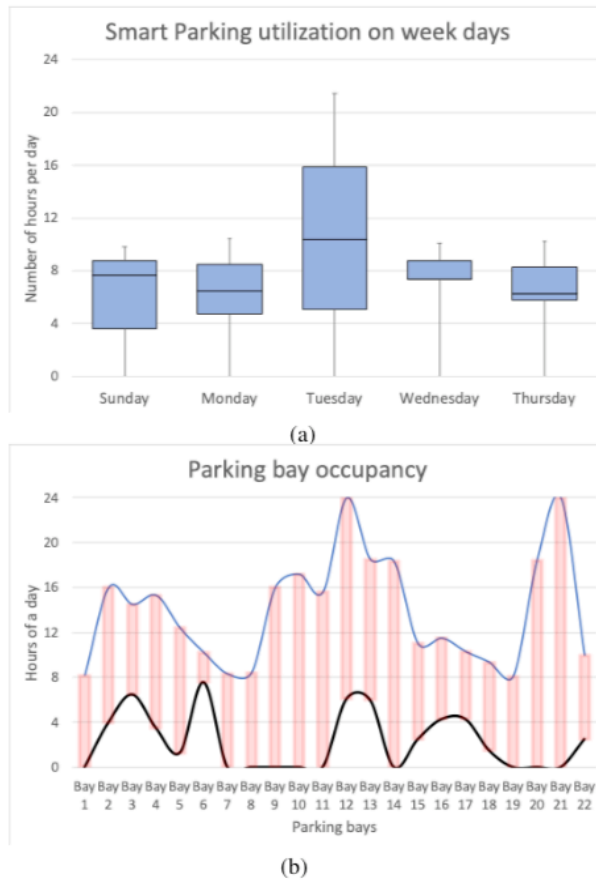

Fig. 5: (a) Smart Parking utilization on weekdays (b) Parking bay max and min occupancy (hours per day)

**Acknowledgments**
This work is supported by the Research and Innovation Center through grant number SSP-18-5-01. This work is also partially supported by the Robotics and Internet of Things Lab. The authors are thankful to Machines-talk for providing wireless parking sensors for this work.